%% file: Manuscript.tex
\documentclass[preprint,3p,times,12pt,authoryear]{elsarticle}
\journal{Preventive Veterinary Medicine}
\usepackage[T1]{fontenc}
\usepackage{graphicx}
\usepackage{caption}
\usepackage[toc,page]{appendix}
\usepackage[section]{placeins}
\usepackage{booktabs}
\usepackage{makecell}
\usepackage{algorithmic}
\usepackage{algorithm2e}
\usepackage{multirow}
\usepackage{float}
\usepackage{pdflscape}
\usepackage{mathtools}
\usepackage{makecell}
\usepackage[symbol]{footmisc}
\usepackage{enumitem} % Import the enumitem package
\usepackage{booktabs}
\usepackage{url} 
\usepackage[flushleft]{threeparttable}
\usepackage{changepage}
\usepackage{lineno}
\usepackage{xcolor}
\usepackage{ragged2e}
\usepackage{adjustbox}
\usepackage{xcolor} % needed for colors
\usepackage[hidelinks]{hyperref} % hide green anoying reference
\usepackage{float}
\usepackage{array}
\newcolumntype{P}[1]{>{\centering\arraybackslash}p{#1}}

\begin{document}

\begin{frontmatter}

 \title{\textbf{A Modelling Assessment of the Impact of Control Measures on Simulated Foot-and-Mouth Disease Spread in Mato Grosso do Sul, Brazil}}

 \author[1]{Nicolas C. Cardenas} 
 \author[2]{Jacqueline Marques de Oliveira}
 \author[2]{Andre de Medeiros C. Lins}
 \author[2]{Fernando Endrigo Ramos Garcia}
 \author[2]{Marcus Vinicius Angelo}
 \author[2]{Robson Campos dos Anjos}
 \author[2]{Fabricio de Lima Weber}
 \author[2]{Frederico Bittencourt Fernandes Maia}
 \author[3]{Vanessa Felipe de Souza}
 \author[1]{Gustavo Machado*}

 \ead{gmachad@ncsu.edu}

% \cortext[cor2]{Corresponding Author.}

% %% Author affiliation
 \affiliation[1]{organization={Department of Population Health and Pathobiology},%Department and Organization
    addressline={North Carolina State University}, 
    city={Raleigh},
    state={NC},
    country={USA}}

% %% Author affiliation
\affiliation[2]{organization={IAGRO, Agência Estadual de Defesa Sanitária Animal e Vegetal},
   city={Campo Grande},
   state={MS},
   country={Brazil}}

\affiliation[3]{organization={Embrapa Gado de Corte},
    city={Campo Grande},    
    state={MS},
    country={Brazil}}
  
\begin{abstract}

This study simulated the introduction of Foot-and-mouth disease (FMD) into Mato Grosso do Sul, Brazil, to evaluate the effectiveness of outbreak control strategies.
Our susceptible-exposed-infected-recovered model generated a range of outbreak sizes across the state. These outbreaks were used to model control actions across six scenarios: high vaccination, two variations of moderate depopulation combined with vaccination, high depopulation with limited vaccination, and moderate and high depopulation alone. Our results showed that relying solely on high vaccination was the least effective approach; it controlled only 2.22\% of outbreaks and resulted in the highest number of infected farms and the longest control duration. Mixed strategies, busing, moderate depopulation, and vaccination controlled approximately 91\% of outbreaks. The use of moderate depopulation alone controlled 
96.60\% of outbreaks, and it was 14-15 days faster than the mixed approaches. The most effective strategy combined the highest depopulation capacity with limited vaccination, controlling 100\% of outbreaks and producing the shortest control duration. The number of vaccinated animals ranged from 211,002 under the optimal strategy to 596,530 when the control strategy included only vaccination. We demonstrated that vaccination alone was insufficient to eliminate outbreaks, and that depopulation and vaccination strategies would be required to stamp out future FMD introduction in Mato Grosso do Sul (MS). The success of such strategy would eliminate between 90\% to 100\% of outbreaks in 10 to 15 days and reduce the number of infected farms by 10 to 13.

\end{abstract}


\end{frontmatter}

\section*{Introduction}

Foot-and-mouth disease (FMD) is a highly contagious viral disease affecting cloven-hoofed animals \citep{brown_airborne_2022}. It is caused by the FMD virus (FMDV; genus \textit{Aphthovirus}, family \textit{Picornaviridae}). The disease poses a persistent threat to livestock production systems worldwide, with significant implications for both commercial and smallholder farming \citep{knight-jones_economic_2013}. Following their introduction in the 1960s, vaccines became a cornerstone of FMD control strategies \citep{sutmoller_control_2003}. In the early 1990s, for instance, mass cattle vaccination campaigns helped eradicate FMD across most of Europe. Upon gaining control, European nations adopted a non-vaccination policy to meet international trade requirements \citep{sutmoller_control_2003}. Similarly, South American countries, including Argentina, Brazil, Chile, Paraguay, and Uruguay, attained FMD-free status through intensive mass vaccination \citep{bergmann_evolving_2005, naranjo_elimination_2013}.

In the Americas, intensive control and eradication efforts have led to most countries being recognized as FMD-free by the Pan American Health Organization \citep{paho_diagnostico_2020}. Despite this progress, the risk of FMD reintroduction remains a major concern, particularly for countries transitioning away from vaccination. In Brazil, the planned cessation of vaccination is expected to increase vulnerability to potential virus incursions. Without the protective buffer conferred by vaccination, the introduction of FMD could spread rapidly through the livestock population, resulting in severe economic losses and trade disruptions \citep{panaftosa_informe_2021, salvarani_what_2025}. Hence, maintaining robust preparedness and response capacities is essential. Establishing a rapid-response system that incorporates quarantine enforcement, prompt culling of infected animals, and immediate containment measures will be critical to limiting disease spread and mitigating potential impacts \citep{salvarani_what_2025}.

The economic consequences of FMD outbreaks are well-documented and can be devastating. In Australia, estimated costs range from AUD \$60 million to \$373 million per outbreak. Similarly, during the 2001 outbreaks in the United Kingdom and the Netherlands, more than six million animals were culled, resulting in direct losses of approximately €3.2 billion, with additional impacts on industries such as tourism estimated at between €2.7 and €3.2 billion. In South America, Argentina reported over 2,100 outbreaks, while Uruguay experienced losses of approximately \$730 million and a 1.9\% reduction in gross domestic product from 2001 to 2003 \citep{iriarte_main_2023, panaftosa_informe_2021, perez_control_2004}. In Brazil, previous assessments estimated potential economic losses between \$132 million and \$271 million, depending on the severity of trade restrictions \citep{de_menezes_potential_2023}. More recently, the 2017--2018 FMD outbreaks in Colombia affected 41 farms, underscoring the continued threat of reintroduction in the region \citep{iriarte_main_2023, panaftosa_informe_2021, perez_control_2004}.

In this context, mathematical models have been developed to simulate the spread of FMD within susceptible animal populations, helping to estimate outbreak sizes, identify high-risk regions, evaluate the effectiveness of control measures, and determine the number of vaccines required to respond to outbreaks in FMD-free countries \citep{cardenas_footandmouth_2025}. These mechanistic models also aid in describing potential outbreak dynamics, estimating economic impacts, and supporting science-based decision-making, ultimately enhancing preparedness and resilience in non-endemic settings (\cite{bjornham_multilevel_2020, cardenas_footandmouth_2025, cardenas_integrating_2025, kirkeby_practical_2021, pesciaroli_modelling_2025, sanson_evaluating_2017}).

This study focuses on Mato Grosso do Sul (MS), a Brazilian state with a livestock population of 20,917,525 animals, including cattle, pigs, sheep, and goats \citep{IBGE_PPM_31810}, which shares extensive borders with Bolivia and Paraguay. In this study, we develop a stochastic, mechanistic model to simulate the spread of FMD within MS and assess the effectiveness of various control strategies across six intervention scenarios. Specifically, we evaluated outbreak duration, the number of culled animals, and the number of vaccinated animals under each strategy.

\section*{Methodology}
\subsection*{Data}\label{data}
\subsubsection*{Population Data}\label{pop_data}
We used data from the official veterinary service of Mato Grosso do Sul (MS), provided by the Agência Estadual de Defesa Sanitária Animal e Vegetal, covering the period from November 30, 2022, to November 30, 2023 \citep{iagro_iagro_nodate}. The dataset covers 103,284 registered farms across 79 municipalities, including 62,894 with bovines, 13,002 with swine, and 7,259 with sheep and goats. Notably, 15,776 farms host more than one species. The spatial distribution of the farms is depicted in Figure \ref{fig:Fig_MS_map}. In addition, the spatial distribution of the farms by species is depicted in Supplementary Material Figure S1, while the distribution quantity of animals by 10 km by host species is presented in Supplementary Material Figure S2. 

\begin{figure*}[!htb]
 \centering
 \includegraphics[width=1\columnwidth]{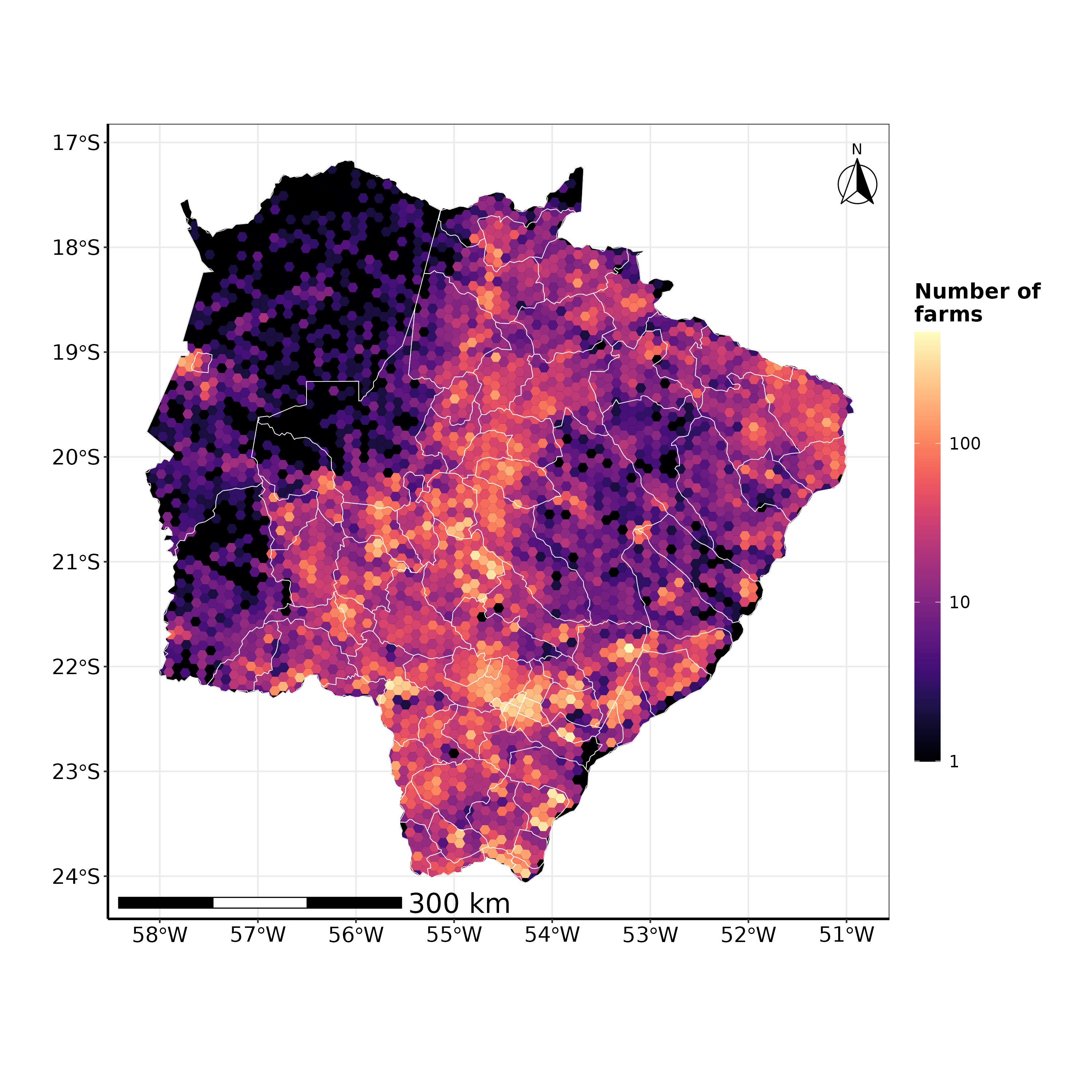}
 \caption{\textbf{Farm density represented in a 10 km$^2$ hexagon grid}. The hexagon intensity color represents the number of farms allocated; note that the color scale is log10. White lines represent the municipality's division of the state of Mato Grosso do Sul, Brazil.}
 \label{fig:Fig_MS_map}
\end{figure*}

\subsubsection*{Events Data}\label{event_data}
The dataset captures four primary event types related to animal entry and exit at the farm level: farm-to-farm movements, farm-to-slaughterhouse movements, births, and deaths, provided by \citep{iagro_website}. Farm-to-farm movements accounted for 267,158 records, followed by births (74,005), movements to slaughterhouses (50,740), and deaths (46,913). Death events reflect on-farm mortalities due to causes other than slaughter, distinguishing them from movements to slaughterhouses. Birth and death events are self-reported by farm owners and are required for eligibility for compensation in the event of an outbreak \citep{decreto_mt_1260_2017}. The final dataset comprises 428,643 bovine-related events, 7,512 small ruminant events, and 2,661 swine events recorded between November 1, 2022, and November 30, 2023. The weekly distribution of these events is presented in Supplementary Material Figure S3.

\subsection*{SEIR Model Description}
\subsubsection*{Model Description and Formulation}\label{model_descript_and_formulation}
A multi-host, single-pathogen model, known as MHASpread, was developed by \citep{cespedes_cardenas_mhaspread_2024, cardenas_footandmouth_2025, cespedes_cardenas_modeling_2024} to simulate FMD outbreaks. The model is based on a susceptible-exposed-infected-recovered (SEIR) compartmental framework, in which the population, comprising bovine, swine, and small ruminants, is stratified by species and allocated into four epidemiological compartments. Within each farm, individuals are assumed to mix homogeneously, allowing for direct and indirect transmission among species.

The transmission of FMD within farms is modeled using species-specific transmission probabilities that capture differences in susceptibility and infectiousness across host species. These parameters reflect the heterogeneous transmission dynamics observed in FMD. All within-farm disease transmission coefficients are summarized in Table \ref{tab:transmission_beta}. Disease progression within each host species follows latent and infectious period distributions presented in Table \ref{tab:S_periods}, after which infected animals transition to the recovered compartment and are assumed to be immune for the duration of the simulation.

Between-farm transmission is explicitly modeled through recorded animal movements, with animals transitioning between farms while retaining their epidemiological status. Movements are tracked at the compartment level, accounting for the number of susceptible, exposed, infected, and recovered animals at both origin and destination farms. Movements from farms to slaughterhouses are also included, with transported animals permanently removed from the system and therefore no longer contributing to disease transmission. Similarly, on-farm deaths are modeled as random removals from any compartment. Births are incorporated as new susceptible animals entering the farm population.

In addition to movement-based transmission, spatial transmission between farms is modeled using a distance-dependent transmission kernel, which captures local spread through mechanisms such as airborne transmission, shared equipment, personnel, and wildlife-mediated contacts. The probability of transmission decreases with increasing inter-farm distance, with a maximum effective range of 40 km, consistent with previous FMD modeling studies \citep{boender_common_2023, boender_transmission_2010}. Details of this approach are presented in Supplementary material methods and in \citep{cespedes_cardenas_mhaspread_2024}. 

\input{table_trasnmission}

While the progression of the disease through various compartments for each species within the farm population is outlined in Table \ref{tab:S_periods}.

\input{infec_incub_periods}

A detailed description of the model is provided in the Methods section of the Supplementary Material. The MHASpread framework, which simulates FMD spread within and between farms, is available as R and Python packages and can be looked up at \url{https://github.com/machado-lab/MHASPREAD-model}.

\subsection*{FMD Spread and Control Actions}\label{control_actions_model}
\subsubsection*{Initial conditions}\label{init_conditions_model}

We evaluated six sets of control actions. To ensure comparability, initial outbreak scenarios were generated and then used to simulate countermeasures under different control settings. The outbreaks were simulated by selecting a representative sample of 1,035 farms in Mato Grosso do Sul to serve as initial infected farms, using a multistage, stratified random sampling approach stratified by host species and municipality \citep{onyeka_estimation_2015}. Farms were stratified to ensure proportional representation across relevant strata, and the final sample within each stratum was randomly selected. The sample size was determined using Cochran's formula for finite populations, assuming a 50\% prevalence, a 95\% confidence level, and a 1.1\% margin of error.

The initial infection was introduced by seeding five infected bovines into each selected farm, regardless of population size. If no bovines were present, small ruminants were infected instead, followed by swine if neither species was available. For each selected farm, the model was run for 20 days without control measures, resulting in outbreaks ranging from 1 to 84 farms. These initial outbreaks were used to benchmark the performance of the proposed control scenarios (described below). The spatial distribution of sampled farms is shown in Supplementary Figure S4, while Figure \ref{fig:figure_distri_20_days} presents the frequency distribution of infected farms.

\begin{figure}[H]
 \centering
 \includegraphics[width=1\columnwidth]{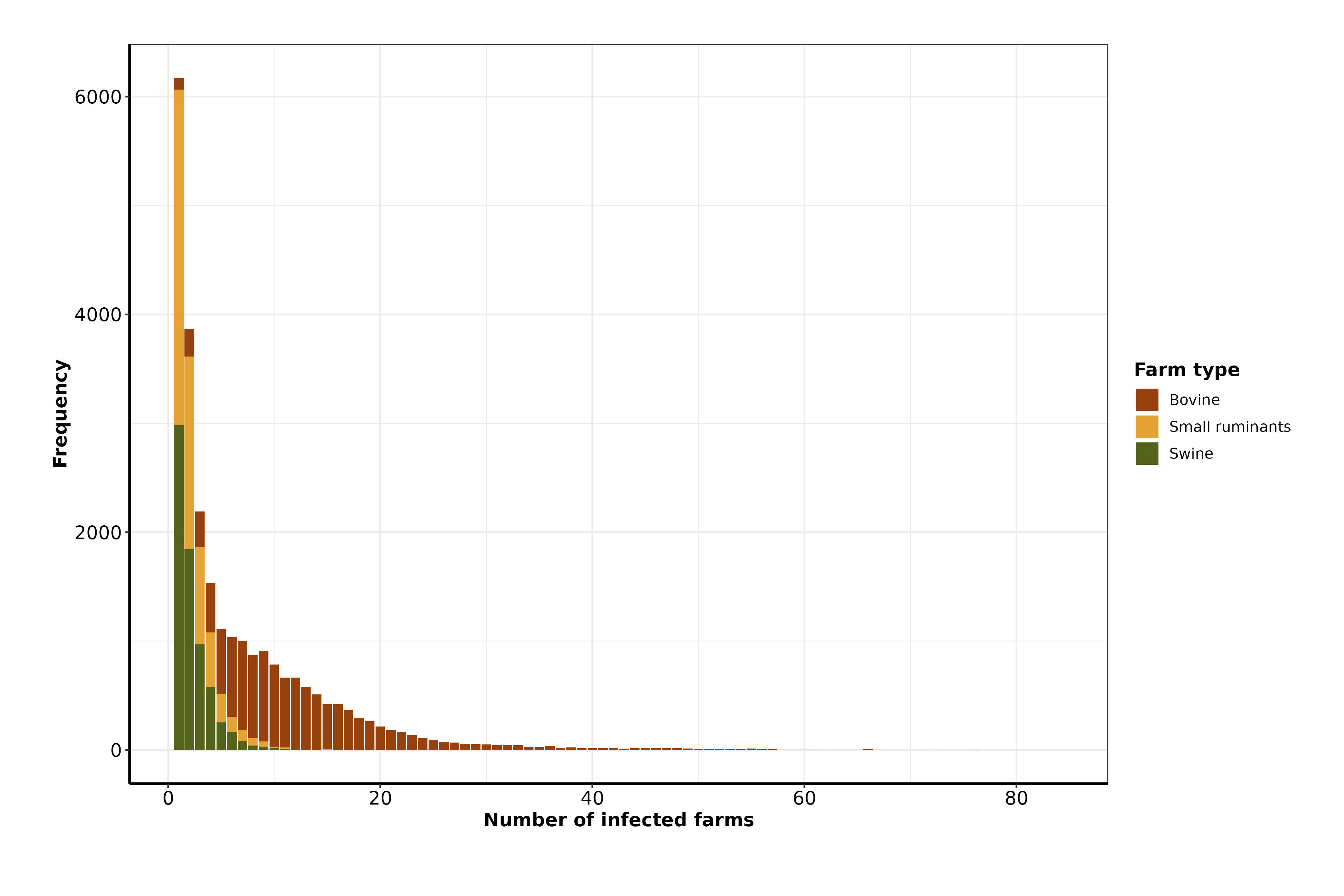}
 \caption{\textbf{Initial conditions.} Distribution of secondary infected farms used in the scenarios for the control strategies section of the model. Color represents the presence of at least one species per farm. For mixed-species farms, classification is hierarchical based on the highest-risk species present (bovine > small ruminants > Swine).}
 \label{fig:figure_distri_20_days}
\end{figure}

\subsubsection*{Scenarios for control strategies}

We outline six distinct control scenarios based on the Brazilian national response plan \citep{mapa_febre_2020}. These strategies were developed in collaboration with \citep{iagro_iagro_nodate} animal health officials and were selected to align with local capacities and the organizational structure of the official veterinary service. 
Each scenario includes establishing three control zones around infected farms and a specific set of countermeasures to contain and eradicate epidemics, including the depopulation rate of infected farms, the emergency vaccination rate, the duration of animal movement standstill, and trace-back settings. The scenario names, rates, and settings of each parameter for each scenario are presented in Table~\ref{tab:model_parameters}.

Table \ref{tab:model_parameters} summarizes the parameterization of the epidemiological model across six control scenarios used in the simulation framework, which differ in the intensity and combination of depopulation and vaccination strategies. All scenarios share common initial conditions, with both the initially infected farm and the initial day of simulation selected at random, and a fixed maximum duration of 120 days for the implementation of control actions. Spatial control measures are standardized across scenarios, with infected, buffer, and surveillance zones defined by radii of 3, 5, and 7 km, respectively. Movement restrictions include a 30-day standstill in the surveillance zone in all scenarios, while additional standstills in infected and buffer zones and the depth of traceback in the contact network vary by scenario, reflecting different levels of tracing intensity. Depopulation strategies range from no depopulation to high capacity daily limits, with some scenarios restricting depopulation exclusively to infected farms and others not applying depopulation at all. Vaccination parameters further differentiate scenarios, including the presence or absence of vaccination, daily vaccination capacity by zone, vaccine efficacy values, species targeted, spatial prioritization of vaccination, and delays to immunity. Collectively, these configurations allow for a systematic comparison of control strategies that vary in operational capacity and intervention mix while maintaining consistent baseline assumptions.

% Table 1 parameter here 
\input{Table1.tex}

We further categorized control scenarios as either “controlled” or “not controlled”. A model simulation was considered “controlled” if no infected farms were present by the end of the simulation. Conversely, a simulation was classified as “not controlled” if the number of infected farms exceeded 400 at any point during the simulation or if control measures extended beyond 100 days while more than one infected farm remained.

\subsubsection*{Performance of control strategies}
To evaluate the performance and differences among the scenarios, we analyzed the percentage of outbreaks successfully controlled under each scenario, the distribution of infected farms at the end of each simulation, the duration of control actions for both controlled and uncontrolled outbreaks, the total number of vaccinated animals, and the total number of depopulated animals at the end of each simulation. The distributions of vaccinated and depopulated animals were further described and compared by assessing frequencies and applying the Kruskal-Wallis rank-sum test to identify significant differences among scenarios. Where significant differences were detected, Dunn's test with Bonferroni corrections was used for multiple pairwise comparisons.

\subsubsection*{Number of vaccinated animals}
We quantified the total number of vaccinated animals in each simulation and summarized these results using the median and interquartile range. For each simulation, the daily number of vaccinated farms was generated by sampling from a predefined, date-specific vaccination rate specified for each scenario (see \ref{tab:model_parameters}). Although MHASpread supports vaccination of multiple species, in this study, we restricted vaccination to bovines and buffaloes. We assumed that the vaccine used was homologous to the simulated FMD virus strain and conferred six months of protection with a single dose before a booster dose was required \citep{ulziibat_comparison_2023}. Furthermore, we assumed that it required 15 days to achieve immunization coverage levels of 0.8, 0.9, or 1.0; at each simulation run, the model randomly sampled one of these coverage levels. Under these assumptions, we calculated the number of animals to be vaccinated rather than the number of vaccine doses administered. Vaccination of cattle was implemented in both infected and buffer zones (Table~\ref{tab:model_parameters}).

Furthermore, in our model, depopulation is prioritized over vaccination, while depopulated farms are excluded from immunization because no animals remain to be vaccinated. Conversely, if animals are vaccinated and subsequently depopulated, they are still recorded as vaccinated even if depopulation occurs in the following days. The control zone area (infected and/or buffer) and the number of farms vaccinated per day were determined according to each scenario (Table \ref{tab:model_parameters}).

\subsection*{Software}
The MHASpread model was initially developed in Python v. 3.8.12 and subsequently analyzed in R v. 4.2.3 (R Core Team, 2025) to create graphics, tables, and maps. In Python, the analysis relied on the following packages: NumPy \citep{harris2020array}, Pandas \citep{mckinney2010data}, and SciPy \citep{virtanen2020scipy}. In R, the following packages were used: sampler \citep{baldassaro_sampler_2019}, tidyverse \citep{wickham_welcome_2019}, sf \citep{pebesma_simple_2018}, doParallel \citep{corporation_doparallel_2022}, and lubridate \citep{grolemund_dates_2011}.

%--------------------------------------------------------------------%
\section*{Results}\label{results}
\subsection*{General results}

We assessed the median number of infected farms at the end of each simulation. The overall median number of infected farms was 51 (interquartile range: 33–79; maximum: 309). Significant differences were observed among scenarios when comparing the distribution of the total number of infected farms at the end of the simulations (Table \ref{tab:disease_control}). Scenario D1V3 differed significantly from D0V4 and D2V2 (p = 0.0032 and p = 0.0013, respectively), demonstrating the benefits of the vaccination strategy combined with moderate depopulation in terms of outbreak containment compared with scenarios with either higher vaccination (D0V4) or different depopulation levels (D2V2). Scenario D3V0 showed significant differences compared to D1V3, D2V0, and D2V2 (p < 0.0001), reflecting the impact of intensive depopulation without vaccination, which reduced both the outbreak duration and the total number of infected farms. Scenario D4V1 was distinct from all other scenarios (p < 0.0001), achieving full outbreak control through a highly targeted combination of rapid depopulation and limited vaccination, resulting in fewer infected farms, shorter control durations, and lower total vaccination requirements. These differences are presented in Supplementary Material Figure S5.

\input{Table2.tex}

\subsection*{Number of vaccinated animals by scenario}
\begin{figure*}[!h]
 \centering
 \includegraphics[width=1\columnwidth]{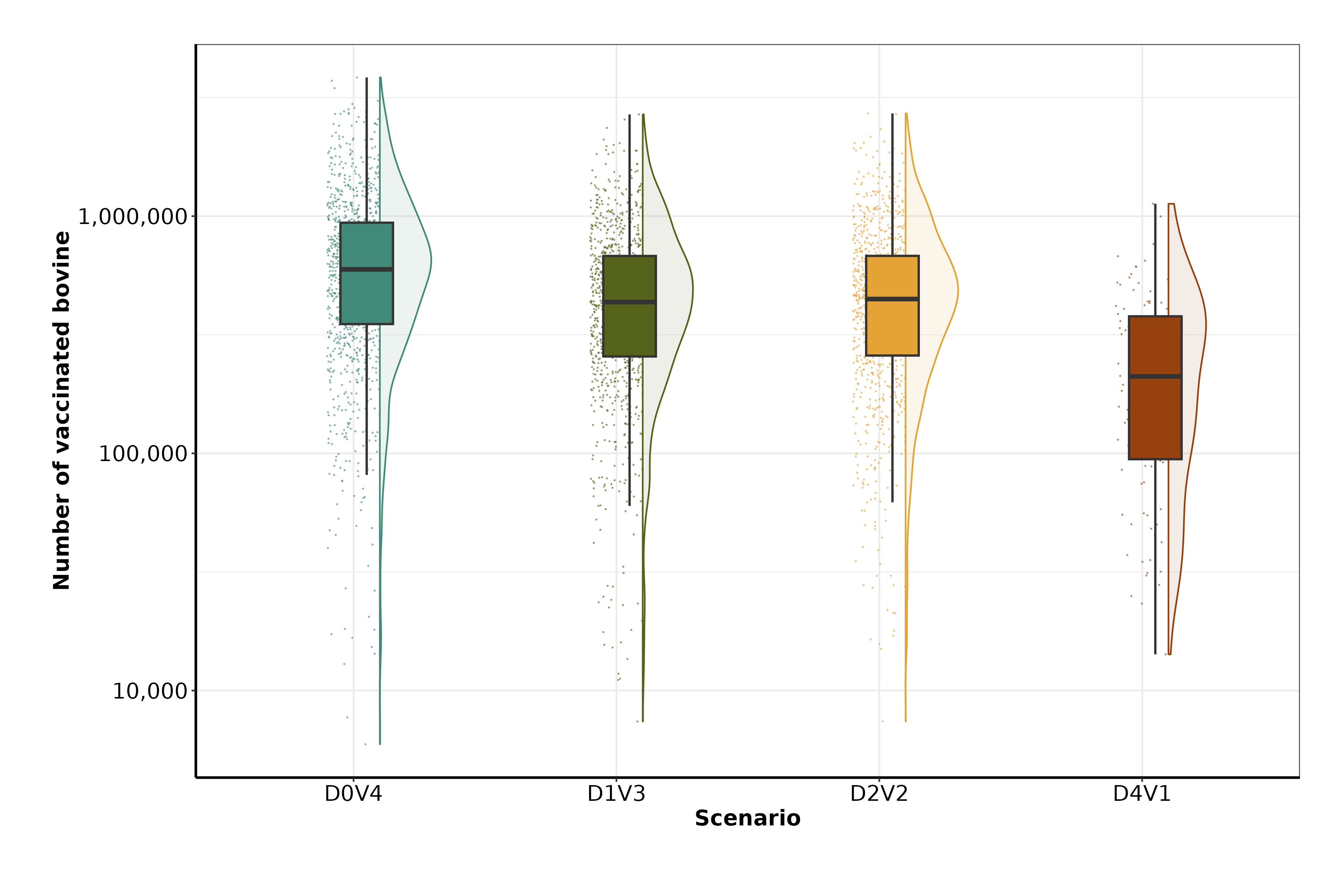}
 \caption{\textbf{Distribution of the total vaccinated bovines. The y-axis denotes the number of vaccinated bovines, while the x-axis represents simulated scenarios.} The dots and shaded area represent the distribution of individual simulations.}
 \label{fig:figure_vaccine}
\end{figure*}

The distribution of vaccinated animals at the end of each simulation by scenario is presented in Figure \ref{fig:figure_vaccine}. The number of vaccinated animals differs significantly across scenarios. Scenario D0V4 showed the highest median number of vaccinated animals at 596,530, which was significantly higher compared to D1V3 and D2V2 (p = 0.0032 and p = 0.0013, respectively). Scenario D2V2 was the second with the most vaccine use, with a median of 447,278 vaccinated bovines, reflecting a moderate reduction from D0V4, and no significant difference compared to D1V3 (p > 0.5). Similarly, scenario D1V3 has a median of 434,240 vaccinated animals. In contrast, scenario D4V1 exhibits the lowest vaccination numbers among all scenarios, with a median of 211,002, showing significant differences from all other scenarios (p < 0.0001). The results of all pair comparisons are presented in the Supplementary Material Figure S6.

\subsection*{Number of depopulated animals}
\begin{figure*}[!h]
 \centering
 \includegraphics[width=1\columnwidth]{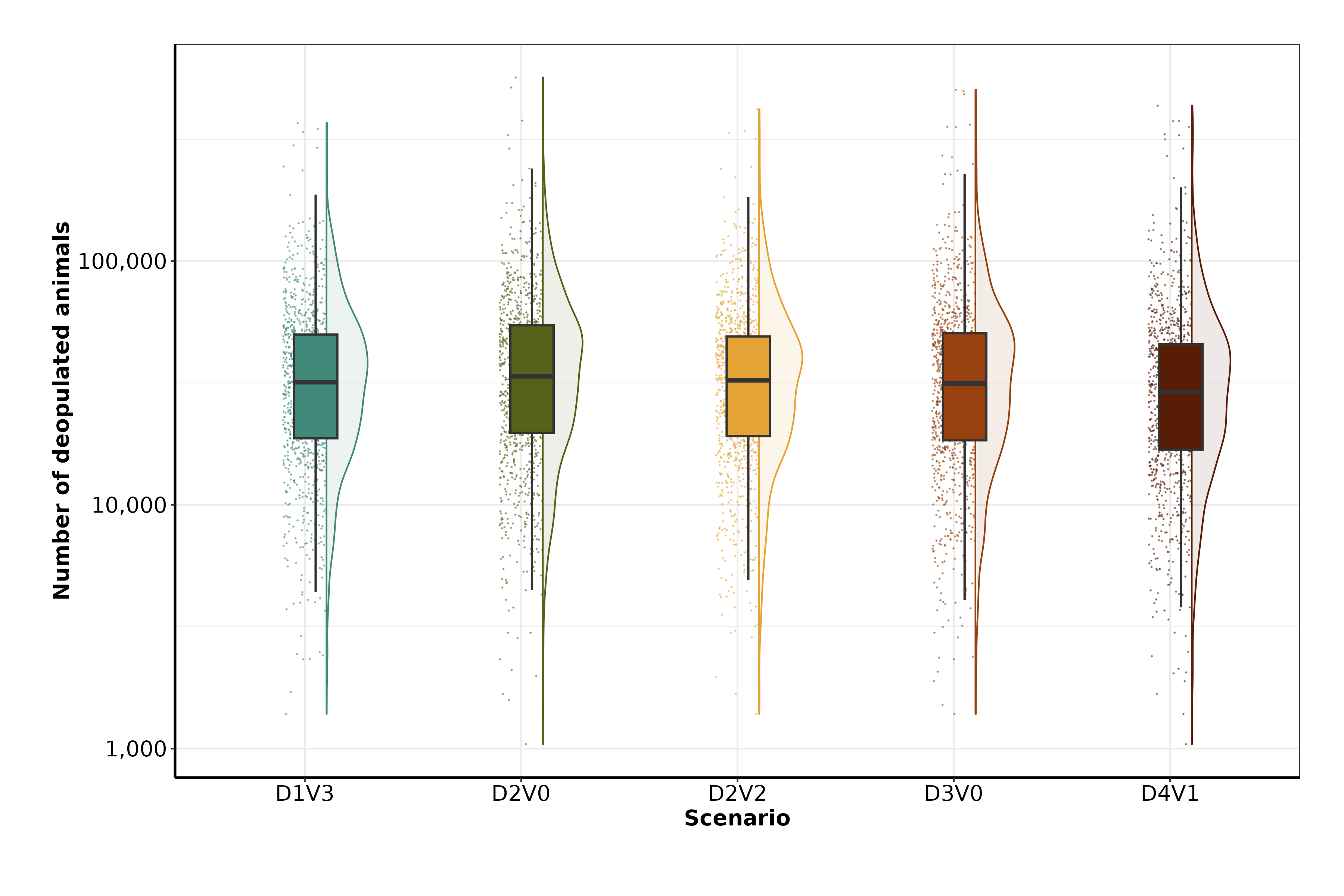}
 \caption{\textbf{Distribution of number of depopulated animals.} The y-axis denotes the number of depopulated animals, while the x-axis represents simulated scenarios. The dots and shaded area represent the distribution of individual simulations.}
 \label{fig:figure_depop}
\end{figure*}

 The distribution of depopulated animals at the end of each simulation is depicted in Figure \ref{fig:figure_depop}, and the complete results are shown in Table \ref{tab:disease_control}. Overall, the number of depopulated animals varied widely across simulations, with a median from 28,976 to 33,097 and ranging from 1,042 to 566,938. The highest number of depopulated animals were D2V0, D3V0, and D2V2, with medians of 33,097 (max: 566,938), 31,467 (max: 504,437), and 30,894 (max: 317,969), respectively. Significant differences were observed between D4V1 to D2V0 (p < 0.0001), D4V1 to D2V2 (p = 0.012), and D4V1 to D1V3 (p = 0.032), as well as between D4V1 and D3V0 (p = 0.044). The results of all pairwise comparisons are provided in Supplementary Material Figure S6.

%\clearpage
\section*{Discussion}
This study applied the MHASpread model to simulate the potential dissemination of FMD in Mato Grosso do Sul, Brazil, and to assess the effectiveness of the broad array of control strategies included in the Brazilian national response plan \citep{mapa_febre_2020}. We assessed six control scenarios designed to contain and mitigate disease spread under different combinations of depopulation and vaccination. Integrated control strategies combining depopulation and vaccination were consistently the most effective, as they simultaneously reduced the number of infectious animals and limited onward transmission through the development of population immunity.

Scenarios that incorporated both depopulation and vaccination (D1V3, D2V2, and D4V1) achieved the highest outbreak control rates, ranging from 90.52\% to 100\% within a median of 36 (IQR: 22–53) and 6 (IQR: 4–9) days, respectively. This occurs because depopulation rapidly removes infectious and high-risk animals, while vaccination progressively reduces susceptibility among remaining populations \citep{iriarte_main_2023, roche_evaluating_2015}. In contrast, vaccination alone was insufficient to eliminate all outbreaks, as vaccine-induced immunity requires time to develop and does not immediately interrupt transmission during the early phase of an outbreak \citep{iriarte_main_2023, roche_evaluating_2015}.

Scenarios D1V3 and D2V2 required longer control periods, with median durations of 54 days (IQR 36–77) and 36 days (IQR 22–53), respectively, compared with depopulation-only strategies, which achieved control 15 to 40 days faster, while maintaining similarly high outbreak control rates. This is explained by the delay between vaccination and the acquisition of protective immunity, during which residual transmission can continue and prolong control efforts \citep{porphyre_vaccination_2013}. Consequently, vaccination may initially slow outbreak resolution when not paired with sufficiently rapid and targeted depopulation; such events have been described in the Argentina outbreak in 2001 \citep{perez_control_2004}.

By contrast, scenario D4V1 demonstrates that when vaccination is optimally integrated with depopulation, the duration of control actions can be substantially reduced. In this scenario, depopulation effectively decreases early transmission, while limited but well-targeted vaccination reduces susceptibility in surrounding at-risk farms, preventing secondary spread. As a result, D4V1 achieved complete outbreak control with a median duration of six days (IQR: 4–9; maximum: 41 days). These findings indicate that vaccination does not inherently shorten control periods when combined with depopulation; rather, its impact depends on the intensity, timing, and spatial targeting of vaccination, as well as its coordination with depopulation strategies aimed at infected and high-risk farms. This result agrees with similar simulation studies in Bolivia, Brazil, Thailand, and Italy \citep{cardenas_footandmouth_2025, cespedes_cardenas_modeling_2024, wongnak_stochastic_2024, pesciaroli_modelling_2025, bellini_simulating_2025}.

Vaccination alone was largely ineffective, as shown by D0V4, which achieved only 2.22\% outbreak control. Depopulation alone, as in D3V0, achieved high efficacy (99.90\%) with 14 days of control (IQR: 8–22), indicating that depopulation is highly effective; however, it may be operationally demanding \citep{knight-jones_economic_2013, wittwer_economic_2024}. Overall, in our simulation in Mato Grosso do Sul, the duration of control actions was closely linked to the number of depopulated and vaccinated animals. Scenarios relying primarily on depopulation, such as D3V0, required longer intervention periods due to logistical challenges of large-scale culling. In contrast, D4V1, which integrated vaccination, significantly reduced control duration while also decreasing the median number of depopulated animals (28,976; IQR: 16,836–45,627) compared to D3V0 (31,467; IQR: 18,393–50,620). Interestingly, high-intensity depopulation reduces the number of vaccinated animals. In D4V1, a median of 211,002 (IQR: 94,520–378,256) animals were vaccinated, which is 51.4\% fewer than in D1V3 (434,240 animals vaccinated), while the median number of depopulated animals decreased only slightly (4.8\%) compared to D1V3 (30,444 depopulated). Similar results were reported in simulated outbreaks in the central United States and Australia \citep{capon_simulation_2021, mcreynolds_modeling_2014, yadav_epidemiologic_2022}. This is because an effective depopulation from the outset helps prevent secondary infections, reducing the need for additional control zones and ultimately leading to fewer vaccinated animals \citep{backer_vaccination_2012, capon_simulation_2021, cardenas_footandmouth_2025}. Despite the strong performance of depopulation alone, our results suggest that early detection combined with small outbreak size allows depopulation to effectively eliminate outbreaks without additional intervention. However, in cases of large-scale outbreaks, vaccination creates a synergy with depopulation that enables more efficient mitigation and control, as demonstrated in large outbreaks in Argentina and Uruguay \citep{iriarte_main_2023, perez_control_2004}. The MHASpread model was also applied in Bolivia \citep{cardenas_footandmouth_2025}, Rio Grande do Sul (RS)\citep{cespedes_cardenas_modeling_2024}, Brazil, which allows for contrasting some vaccine numbers under comparable modeling settings. In Bolivia, a median of 1,962 animals were vaccinated per day (IQR: 726 to 6,752), with 50 to 90 farms vaccinated per day. RS achieved a similar median of 1,928 animals per day (IQR: 1,562 to 3,567; maximum: 20,740), but with higher per-farm capacity, vaccinating approximately 20 farms per day. In contrast, MS operated at a substantially larger scale, with a median of 9,247 animals vaccinated per day (IQR: 4,702 to 16,770), vaccinating 60 to 100 farms per day. Under these settings, Bolivia controlled approximately 60\% of cases, whereas RS and MS achieved outbreak control rates exceeding 90\%. Across all scenarios, the depopulation rate remained consistent at 1 to 2 farms per day, highlighting that vaccination requirements are strongly influenced by population density and vaccination intensity. Consequently, the number of animals requiring vaccination cannot be reliably determined by direct extrapolation from other regions \citep{cardenas_footandmouth_2025, cespedes_cardenas_modeling_2024, meadows_disentangling_2018}.

\subsection*{Limitations and further remarks}
While the MHASpread model \citep{cespedes_cardenas_mhaspread_2024} provides insights into the dynamics of FMD transmission, but limitations should be acknowledged. First, the model assumes homogeneous mixing of animals within farms, which may oversimplify intra-farm transmission processes. In reality, contact patterns are often structured by production stage, housing, and management practices \citep{IBGE_CensoAgropecuario2017}. Future extensions could incorporate heterogeneous contact structures to better represent within-farm transmission dynamics \citep{10.3389/fvets.2020.527558, craft_infectious_2015}.

The model relies on historical animal movement and population data, which may not fully reflect real-time changes. The absence of historical FMD outbreaks in the study region further limited our ability to calibrate and validate the model \citep{10.1371/journal.pcbi.1007893}. As a result, epidemiological parameters were derived from the literature rather than estimated from observed outbreak data, and independent validation using epidemic trajectories was not performed. The analysis is geographically restricted to a single state. Although the findings are relevant to livestock systems with similar structural characteristics, extrapolation to other regions should be approached with caution, as differences in farm density, production systems, and movement patterns may influence disease dynamics and control effectiveness.

Finally, while no major data gaps were identified, the analysis depends on official registries, and the potential absence of unregistered farms cannot be entirely excluded. Farm population sizes and demographic events are self-reported by producers. In addition, illegal or unreported animal movements could not be incorporated due to data limitations. Nevertheless, this is unlikely to have substantially affected the results, as animal movements require official authorization and are restricted to the number of animals registered in the system. The volume of information is sufficient to provide a representation of the state, as all animal movements require state authorization and are limited to the total number of animals permitted. The authors will continue to develop and refine the MHSpread model, with the objectives of improving computational efficiency, enhancing usability, and extending its applicability to other diseases and analytical tasks, including the economic assessment of outbreaks \citep{cardenas_integrating_2025}, finally future works will assess the number of required vaccine doses over different outbreak sizes.

\section*{Conclusion}
We demonstrated that integrated scenarios combining vaccination and depopulation are significantly more effective than single-method approaches, achieving containment rates between 90.96\% and 100\%. While vaccination alone was ineffective, controlling only 2.22\% of outbreaks, its integration with depopulation substantially improved outcomes. For instance, D4V1 achieved 100\% control with a median of 6 days to contain the outbreak, compared with 14 days in the depopulation-only scenario. Moreover, high-intensity depopulation requires fewer animals to be depopulated, as it prevents the onset of new cases. The likelihood of FMD eradication depends on maintaining adequate vaccine coverage, rapid deployment, and the capacity to depopulate on a daily basis. Under our specific control strategies, the median number of animals that were vaccinated ranged from 211,002 to 596,530, whereas the number of animals subjected to depopulation varied from 28,976 to 33,097. Early detection and swift implementation of control measures remain critical for outbreak containment and minimizing epidemic duration.

\section*{Acknowledgments}

We would like to thank the animal health officials of agência Estadual de Defesa Sanitária Animal e Vegetal, Campo Grande, MS, Brazil (IAGRO) for their constant contributions to discussions regarding the state FMD response plan. 

\section*{Funding statement}
This work was supported by the Machado Laboratory at the Department of Population Health and Pathobiology, College of Veterinary Medicine, North Carolina State University, Raleigh, NC, USA.

\section*{Data Availability Statement}		
The data supporting this study's findings are not publicly available and are protected by confidential agreements; therefore, they are not available.

\bibliographystyle{elsarticle-harv}
\bibliography{references}

\end{document}

\maketitle % Optional: adds the title/author/date at the top

\subsection*{Model formulation and description}
We developed a transmission model that integrates multiple hosts and a single pathogen across different scales to replicate the trajectories of FMD epidemics (Garira, 2018; Cespedes Cardenas and Machado, 2024; Cespedes Cardenas et al., 2024) and enables the simulation of countermeasures. This model resulted in the creation of a software package named \textbf{"MHASpread: A multi-host animal spread stochastic multilevel model"} (version 3.0.0). More detailed information at \url{https://github.com/machado-lab/MHASPREAD-model}. MHASpread enables the explicit definition of species-specific transmission probabilities and the periods during which transmission can occur.

\subsubsection*{Within-farm dynamics}
For the within-farm dynamics, we assumed populations were homogeneously distributed. Species were homogeneously mixed in farms with at least two species, meaning that the probability of contact among species was homogeneous even when species were segregated in barns and/or paddocks (e.g., commercial swine farms are housed in barns with limited chances of direct contact with cattle). The within-farm dynamics comprise mutually exclusive health states (i.e., an individual can be in only one state per discrete time step) for animals of each species (bovines, swine, and small ruminants). Health states (hereafter, ``compartments'') include susceptible ($S$), exposed ($E$), infectious, ($I$), and recovered ($R$), defined as follows:
\begin{itemize}
    \item \textbf{Susceptible:} animals that are not infected and are susceptible to infection.
    \item \textbf{Exposed:} animals that have been exposed but are not yet infectious.
    \item \textbf{Infectious:} infected animals that can successfully transmit infection.
    \item \textbf{Recovered:} animals that have recovered and are no longer susceptible.
\end{itemize}
Our model accounts for births and deaths, which are used to update the population of each farm. The total population is calculated as $N=S+E+I+R$. The number of individuals within each compartment transitions from $S \xrightarrow{\beta} E$, $E \xrightarrow{1/\sigma} I$, $I \xrightarrow{1/\gamma} R$ according to the following equations:
% Equations 1-4
\begin{align}
\frac{dS_i(t)}{dt} &= u_i(t) - v_i(t) - \frac{S_i(t)I_i(t)}{N_i}  \\
\frac{dE_i(t)}{dt} &= \frac{S_i(t)I_i(t)}{N_i} - v_i(t)E_i(t) - \frac{1}{\sigma}E_i(t) \\
\frac{dI_i(t)}{dt} &= \frac{1}{\sigma}E_i(t) - \frac{1}{\gamma}I_i(t) - v_i(t)I_i(t) \\
\frac{dR_i(t)}{dt} &= \frac{1}{\gamma}I_i(t) - v_i(t)  
\end{align}
Transmission depends on infected and susceptible host species, as reflected by the species-specific FMD transmission coefficient ($\beta$) (Supplementary Material Table S1).

% Table S1
\begin{table}[h]
\centering
\caption{The distribution of each host-to-host transmission coefficient ($\beta$) per animal$^{-1}$ day$^{-1}$.}
\label{tab:S1}
\begin{tabular}{lp{3cm}p{4cm}p{6cm}} % adjust widths as needed
\toprule
\textbf{Infected species} & \textbf{Susceptible species} & \textbf{Transmission coefficient ($\beta$)} & \textbf{Reference} \\
& & \textbf{shape and distribution (min, mode, max)} & \\
\midrule
Bovine & Bovine & PERT (0.18, 0.24, 0.56) & Calculated from the 2000-2001 FMD outbreaks in the state of Rio Grande do Sul (da Costa et al., 2022) \\
Bovine & Swine & PERT (0.18, 0.24, 0.56) & Assumed \\
Bovine & Small ruminants & PERT (0.18, 0.24, 0.56) & Assumed \\
Swine & Bovine & PERT (3.7, 6.14, 10.06) & Assumed (Eblé et al., 2006) \\
Swine & Swine & PERT (3.7, 6.14, 10.06) & (Eblé et al., 2006) \\
Swine & Small ruminants & PERT (3.7, 6.14, 10.06) & Assumed (Eblé et al., 2006) \\
Small ruminants & Bovine & PERT (0.044, 0.105, 0.253) & Assumed (Orsel et al., 2007) \\
Small ruminants & Swine & PERT (0.006, 0.024, 0.09) & (Goris et al., 2009) \\
Small ruminants & Small ruminants & PERT (0.044, 0.105, 0.253) & (Orsel et al., 2007) \\
\bottomrule
\end{tabular}
\end{table}

The Mato Grosso do Sul database does not consistently record the number of animal births and deaths (excluding those sent to slaughterhouses). Despite this limitation, our model can incorporate these births and deaths whenever data are available. Here, births are represented by the number of animals born alive $u_i(t)$ that enter the $S$ compartment on the farm $i$ at time $t$ according to the day-to-day records; similarly, $v_i(t)$ represents the exit of the animals from any compartment due to death at time $t$. The transition from $E$ to $I$ is driven by $1/\sigma$, and the transition from $I$ to $R$ is driven by $1/\gamma$; these values are drawn from species-specific distributions derived from the literature (Supplementary Material Table S2).

% Table S2
\begin{table}[h]
\centering
\caption{The within-farm distribution of latent and infectious periods for FMD in each species.}
\label{tab:S2}
\begin{tabular}{lp{2.5cm}p{5cm}p{5cm}} % adjust widths as needed
\toprule
\textbf{FMD parameter} & \textbf{Species} & \textbf{Mean, median (25\textsuperscript{th}, 75\textsuperscript{th} percentile) in days} & \textbf{Reference} \\
\midrule
Latent period, $\sigma$ & Bovine & 3.6, 3 (2, 5) & (Mardones et al., 2010) \\
 & Swine & 3.1, 2 (2, 4) & (Mardones et al., 2010) \\
 & Small ruminants & 4.8, 5 (3, 6) & (Mardones et al., 2010) \\
\midrule
Infectious period, $\gamma$ & Bovine & 4.4, 4 (3, 6) & (Mardones et al., 2010) \\
 & Swine & 5.7, 5 (5, 6) & (Mardones et al., 2010) \\
 & Small ruminants & 3.3, 3 (2, 4) & (Mardones et al., 2010) \\
\bottomrule
\multicolumn{4}{l}{\textit{Note: The time unit is days.}}
\end{tabular}
\end{table}

\subsubsection*{Kernel transmission dynamics}
Spatial transmission encompasses a range of mechanisms, including airborne transmission, animal contact across fence lines, and equipment sharing between farms (Boender et al., 2010; Boender and Hagenaars, 2023). Local spread was modeled using a spatial transmission kernel, in which the likelihood of transmission decreased with distance. The probability $P_{E}$ at time $t$ describes the likelihood that a farm becomes exposed and is calculated as follows:
% Equation 5
\begin{equation}
P_{E_j}(t) = 1 - \prod_{i} \left( 1 - \frac{I_i(t)}{N_i} \phi e^{-\alpha d_{ij}} \right) 
\end{equation}
where $j$ represents the uninfected population and $\alpha d_{ij}$ represents the distance between farm $j$ and infected farm $i$, with a maximum of 40 km (Cespedes Cardenas et al., 2024). Given the extensive literature on distance-based FMD dissemination and a previous comprehensive mathematical simulation study (Björnham et al., 2020), distances above 40 km were not considered. Here, $1 - \frac{I_i(t)}{N_i} \phi e^{-\alpha d_{ij}}$ represents the probability of transmission between farms $i$ and $j$ scaled by infection prevalence of farm $i$, $\frac{I_i}{N_i}$, given the distance between the farms in kilometers. The parameters $\phi$ and $\alpha$ control the shape of the transmission kernel; $\phi = 0.044$, which is the probability of transmission when $d_{ij} = 0$, and $\alpha = 0.6$ control the steepness with which the probability declines with distance (Boender et al., 2010; Boender and Hagenaars, 2023; Cespedes Cardenas et al., 2024).

\subsection*{FMD spread and control actions}
We initially simulated a silent spread over 20 days, generating a broad spectrum of outbreak scenarios before implementing any control measures, as shown in Figure 2. The FMD control scenarios include the following measures: i) depopulation of infected farms, ii) emergency vaccination of farms within the infected and buffer zones, iii) a 30-day standstill on animal movement, and iv) the establishment of three distinct control zones around infected farms: a 3 km infected zone, a 7 km buffer zone, and a 15 km surveillance zone (Supplementary Material Figure S3).

The depopulation of infected farms involves destroying all animals on farms within the infected zone(s), with priority given to farms with larger animal populations. Once depopulated, these farms are excluded from the simulation. When the daily depopulation capacity cannot accommodate all infected farms in a single day, the remaining farms are scheduled for depopulation the following day or as soon as possible, subject to the capacity limits of each simulated scenario (Table 3).

\subsubsection*{Infected farms detection}

Active detection was modeled as a stochastic inspection process applied to farms under surveillance. At each iteration $i$, detection could only occur among farms that simultaneously met three conditions: (i) they were infected, (ii) they were included in the surveillance zone, and (iii) they had not been detected in previous iterations. This restriction ensures that detection represents newly identified infected farms and avoids double-counting.

Let:
\begin{itemize}
    \item $P_i$ denote the total number of farms under surveillance at iteration $i$,
    \item $I_i$ denote the number of infected farms within the surveillance population at iteration $i$,
    \item $U_i$ denote the number of infected farms under surveillance that had not yet been detected,
    \item $k_i = \max\left(1, \lfloor P_i / 3 \rfloor \right)$ denote the number of farms inspected during iteration $i$.
\end{itemize}

The number of farms inspected was defined as one-third of the surveillance population, reflecting limited inspection capacity while ensuring that at least one farm was inspected whenever surveillance was active. If either $P_i = 0$ or $I_i = 0$, no detection was possible, and the detection process was skipped for that iteration.

Detection was modeled as a sampling-without-replacement process. Specifically, the number of infected farms identified during iteration $i$, denoted by $E_i$, was drawn from a hypergeometric distribution:

\begin{equation}
E_i \sim \text{Hypergeometric}(M = P_i,\; n = I_i,\; N = k_i),
\end{equation}

where $M$ is the total number of farms under surveillance, $n$ is the number of infected farms within that population, and $N$ is the number of farms inspected. This formulation explicitly accounts for the dependence between inspected farms and the finite size of the surveillance population, which is appropriate for structured livestock systems.

To represent imperfect diagnostic performance, the initially detected farms were further filtered using a binomial process with test sensitivity $s$:

\begin{equation}
E_i^{*} \sim \text{Binomial}(E_i,\; s).
\end{equation}

This step captures the probability that an infected farm undergoing inspection yields a positive diagnostic result. When $s = 1$, detection is assumed to be perfect, while lower values represent reduced sensitivity due to subclinical infection, sampling error, or diagnostic limitations.

The final number of newly detected farms was constrained such that

\begin{equation}
E_i^{*} \leq U_i,
\end{equation}

ensuring that detection could not exceed the number of eligible undetected infected farms. If $E_i^{*} = 0$, no new farms were added to the detected set. Otherwise, $E_i^{*}$ farms were randomly sampled from the set of undetected infected farms under surveillance and added to the list of detected farms.

\subsubsection*{Vaccination}

Vaccination was applied exclusively to bovine populations. Bovine farms located within the infected and buffer zones received emergency vaccination 15 days after the establishment of the control zone. Due to logistical capacity constraints, farms that could not be vaccinated within a single day were vaccinated on subsequent days.

The number of farms vaccinated per day varied according to the simulated scenario 
(Table 2). Vaccination was administered to farms within the infected and/or buffer zones according to predefined prioritization criteria, prioritizing those with the largest animal populations. Once enrolled, each farm was assigned a specific vaccination start day, and animals from the $S$, $E$, $I$, and $R$ compartments were progressively transferred to the $V$ (vaccinated) compartment over successive time steps.

At each time step, the number of animals attempted for vaccination on a given farm was computed as a fixed daily fraction of the total unvaccinated population:

% Equation 6
\begin{equation}
    a_t = \left\lfloor (S + E + I + R) \cdot \frac{1}{v_t} \right\rfloor
\end{equation}

\noindent where $v_t = 15$ days represents the duration over which the full unvaccinated 
herd is expected to be processed, and $\lfloor \cdot \rfloor$ denotes the floor function. 
The attempt $a_t$ was further capped to prevent overshooting the target coverage:

\begin{equation}
    a_t = \min\!\left(a_t,\; \max\!\left(V^* - V,\; 0\right)\right)
\end{equation}

\noindent where $V^* = \lfloor N \cdot \tau \rfloor$ is the target number of vaccinated 
animals, $N$ is the total herd size, and $\tau$ is the target vaccination coverage 
(set to 100\% in the present study). The number of animals that successfully acquired 
vaccine-induced immunity was then drawn from a binomial distribution:

\begin{equation}
    X \sim \mathrm{Binomial}(a_t,\; v_e)
\end{equation}

\noindent where $v_e$ denotes the vaccine efficacy, representing the probability that 
a vaccinated animal develops protective immunity. The $X$ successful animals were 
subsequently allocated across the $S$, $E$, $I$, and $R$ compartments proportionally 
to their relative sizes, using a sequential hypergeometric sampling procedure. 
Specifically, for each compartment $k$ in turn, the number of animals drawn was 
sampled as:

\begin{equation}
    d_k \sim \mathrm{Hypergeometric}(C_k,\; U - C_k,\; r_k)
\end{equation}

\noindent where $C_k$ is the count of animals in compartment $k$, $U$ is the total 
unvaccinated pool at that step, and $r_k$ is the number of successful vaccinations 
remaining to be allocated. This procedure ensures that animals are removed from each 
compartment in proportion to its size, without replacement. All allocated animals were 
transferred to the $V$ compartment. A farm was marked as fully vaccinated and excluded 
from subsequent processing once the vaccinated population reached the target coverage 
$\tau$ for all enabled species.

\subsubsection*{Traceability and movement standstill}
We employed contact tracing to identify farms with direct links to infected farms within the past 30 days and subjected these farms to surveillance. Positive farms identified through traceback were categorized as detected infected farms, triggering the application of the same criteria to control zones. In addition, a 30-day restriction on animal movement was enforced across all three control zones, prohibiting both incoming and outgoing movements. The control zones remained in place, and the movement standstill was maintained until depopulation efforts were fully completed.

\clearpage % Start figures on a new page

\subsection*{Supplementary Figures}
%%%%%%%%%%%%% Figure 1     %%%%%%%%%%%%
\begin{figure*}[!htb]
 \centering
 \includegraphics[width=1\columnwidth]{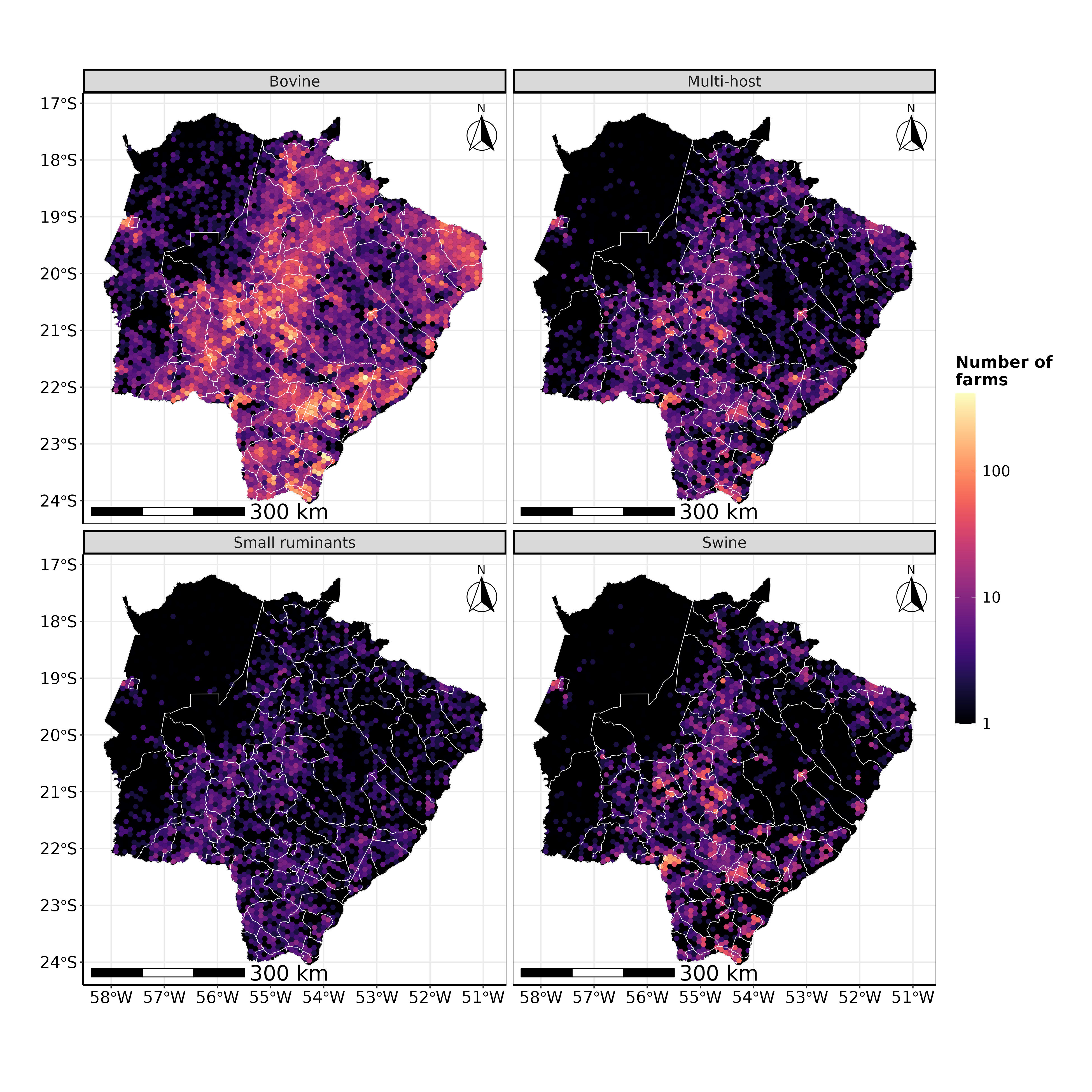}
 \captionsetup{labelformat=empty} % removes "Figure X:"
 \caption{\textbf{Supplementary Figure S1.} \textbf{Farms density population by species.} The size of each hexagon grid is 10 km$^2$. The hexagon color represents the number of farms allocated (log10 scale). White lines represent the municipalities' division of Mato Grosso do Sul, Brazil. Each panel shows farms for each species and farms with more than one species.}
 \label{fig:supp_figure_s1}
\end{figure*}

%%%%%%%%%%%%% Figure 2    %%%%%%%%%%%%

\begin{figure*}[!htb]
 \centering
 \includegraphics[width=1\columnwidth]{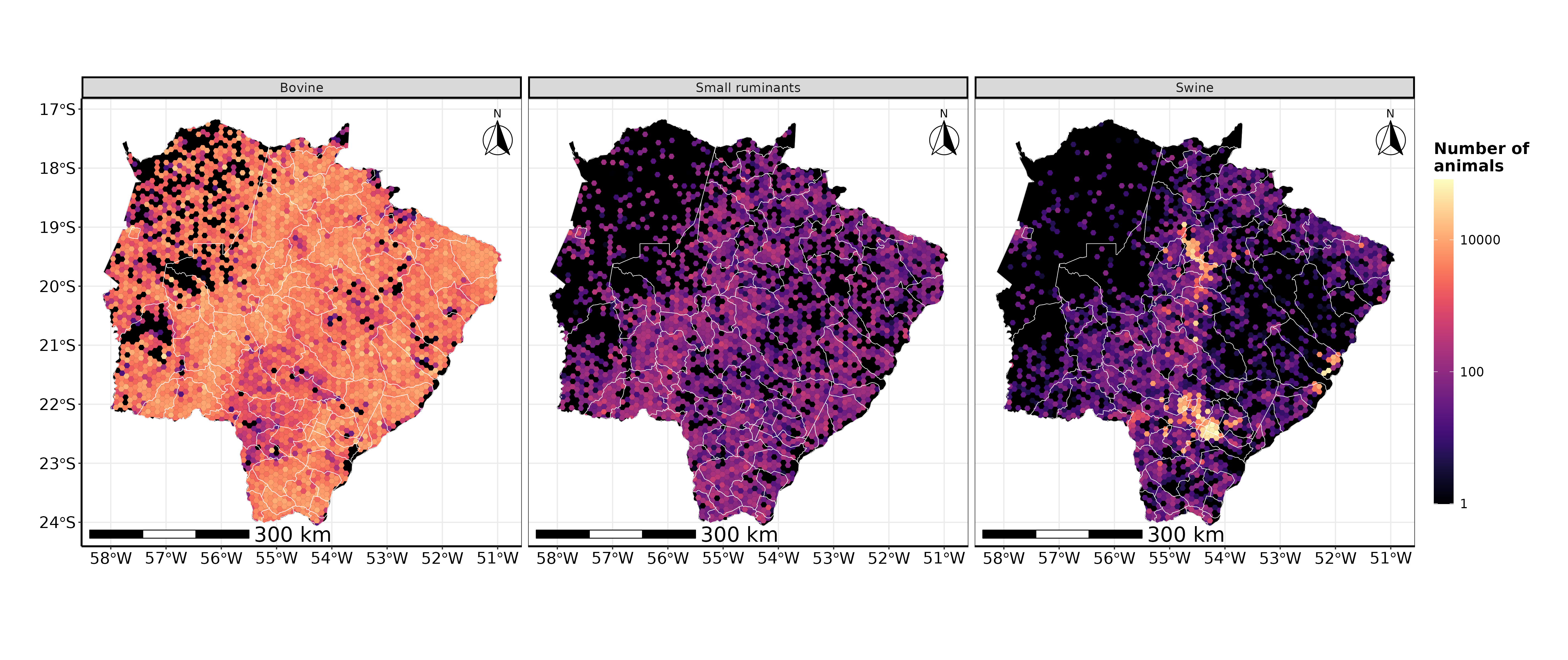}
 \captionsetup{labelformat=empty} % removes "Figure X:"
 \caption{\textbf{Supplementary Figure S2.} \textbf{Animal density population by species.} The size of each hexagon grid is 10 km$^2$. The hexagon color indicates the number of animals allocated; note that the color is on a log10 scale. White lines represent the municipalities' division of Mato Grosso do Sul, Brazil. Each panel represents a specific host species.}
 \label{fig:supp_figure_s2}
\end{figure*}

%%%%%%%%%%%%% Figure 3    %%%%%%%%%%%%

\begin{figure*}[!htb]
 \centering
 \includegraphics[width=1\columnwidth]{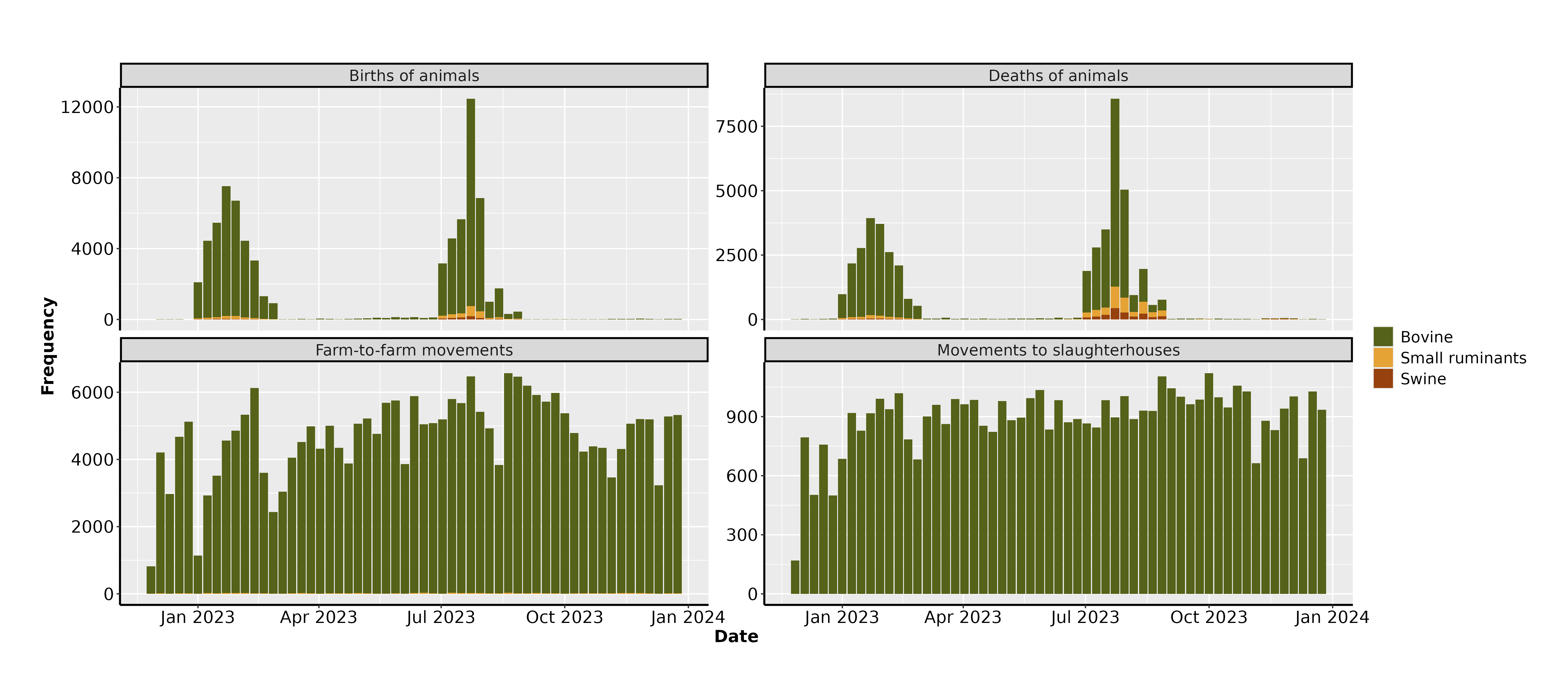}
 \captionsetup{labelformat=empty} % removes "Figure X:"
 \caption{\textbf{Supplementary Figure S3.} \textbf{Frequency of events by species.} The color represents the host species. Each panel represents a specific event.}
 \label{fig:supp_figure_s3}
\end{figure*}

%%%%%%%%%%%%% Figure 4    %%%%%%%%%%%%

\begin{figure*}[!htb]
 \centering
 \includegraphics[width=1\columnwidth]{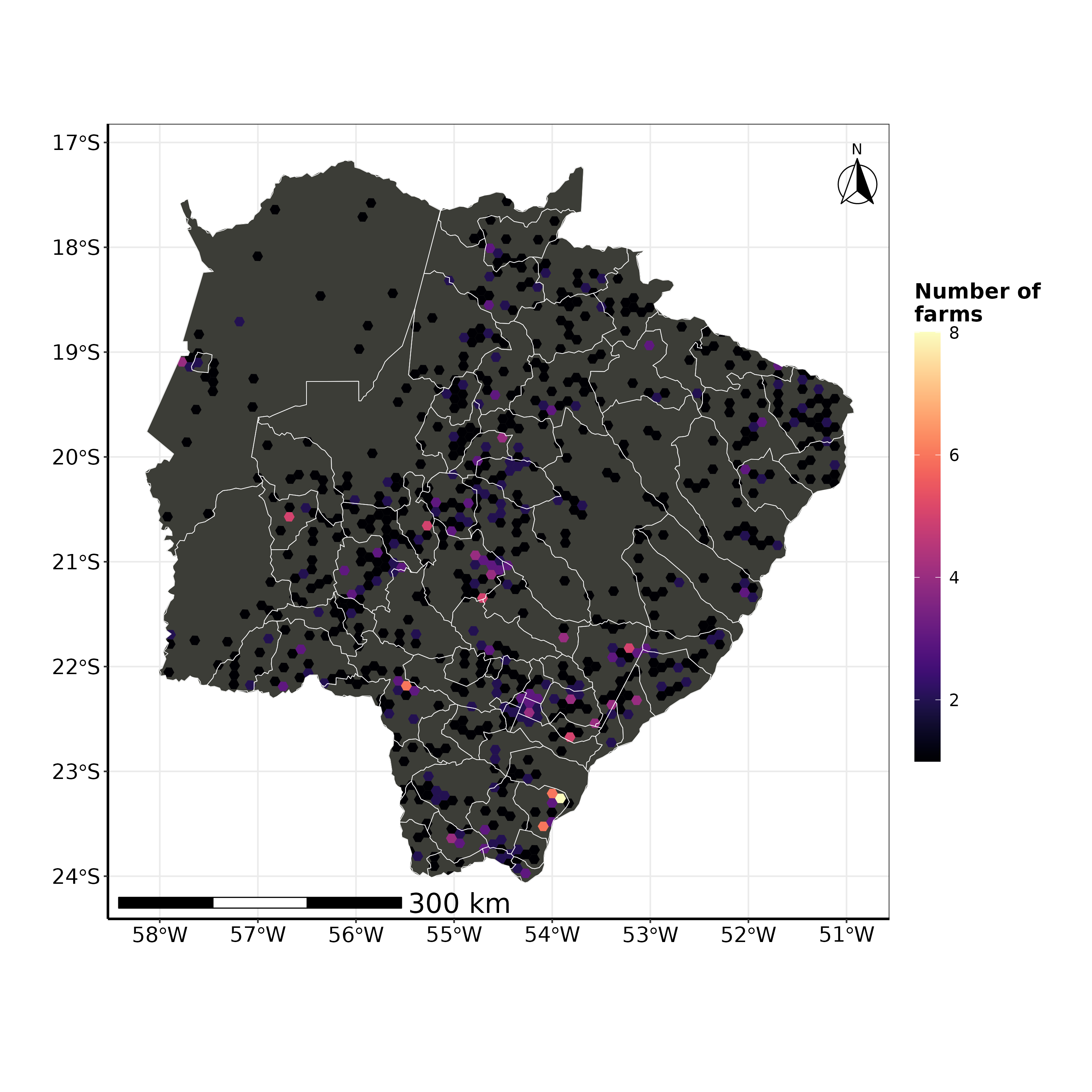}
 \captionsetup{labelformat=empty} % removes "Figure X:"
 \caption{\textbf{Supplementary Figure S4.} \textbf{The grid density of the sampled farms to seed initial outbreaks.} A 10 km$^2$ hexagon grid was projected onto the map, and the density was represented as the number of farms allocated in each grid cell. The gray lines represent the political municipal division of Mato Grosso do Sul, Brazil.}
 \label{fig:supp_figure_s4}
\end{figure*}

%%%%%%%%%%%%% Figure 5    %%%%%%%%%%%%
\begin{figure*}[!htb]
 \centering
 \includegraphics[width=1\columnwidth]{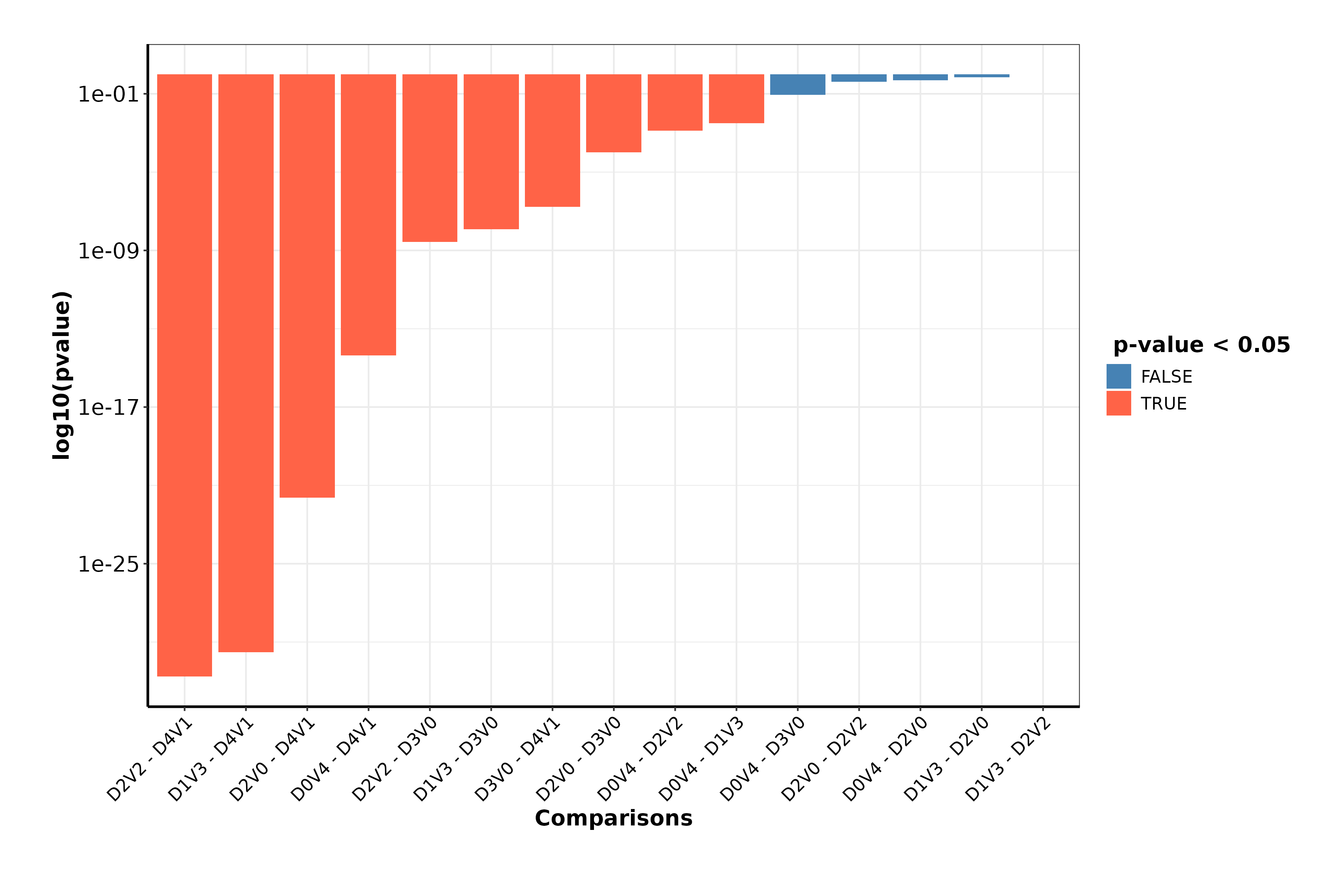}
 \captionsetup{labelformat=empty} % removes "Figure X:"
 \caption{\textbf{Supplementary Figure S5.} \textbf{Results of Dunn's Test p-values for pairwise comparisons between different scenarios for the total number of infected farms.} The y-axis represents the negative log-transformed p-values ($\text{-log}_{10}(\text{p-value})$), with red indicating lower p-values (higher significance) with p-value $<$ 0.05.}
 \label{fig:supp_figure_s5}
\end{figure*}

%%%%%%%%%%%%% Figure 6    %%%%%%%%%%%%
\begin{figure*}[!htb]
 \centering
 \includegraphics[width=1\columnwidth]{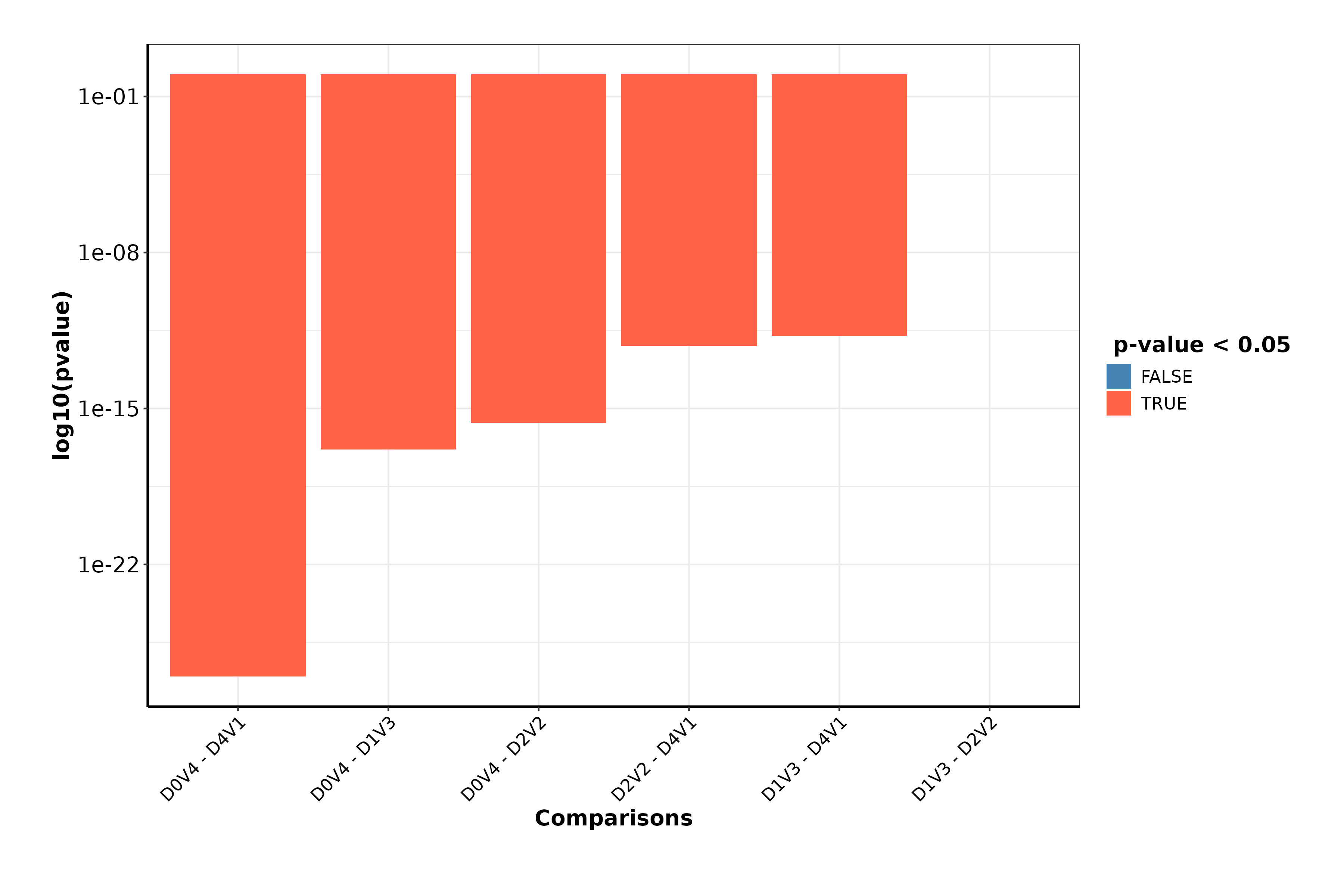}
 \captionsetup{labelformat=empty} % removes "Figure X:"
 \caption{\textbf{Supplementary Figure S6.} \textbf{Results of Dunn's Test p-values for pairwise comparisons between different scenarios for the total number of vaccinated bovines.} The y-axis represents the negative log-transformed p-values ($\text{-log}_{10}(\text{p-value})$), with red indicating lower p-values (higher significance) with p-value $<$ 0.05.}
 \label{fig:supp_figure_s6}
\end{figure*}

%%%%%%%%%%%%% Figure 7   %%%%%%%%%%%%
\begin{figure*}[!htb]
 \centering
 \includegraphics[width=1\columnwidth]{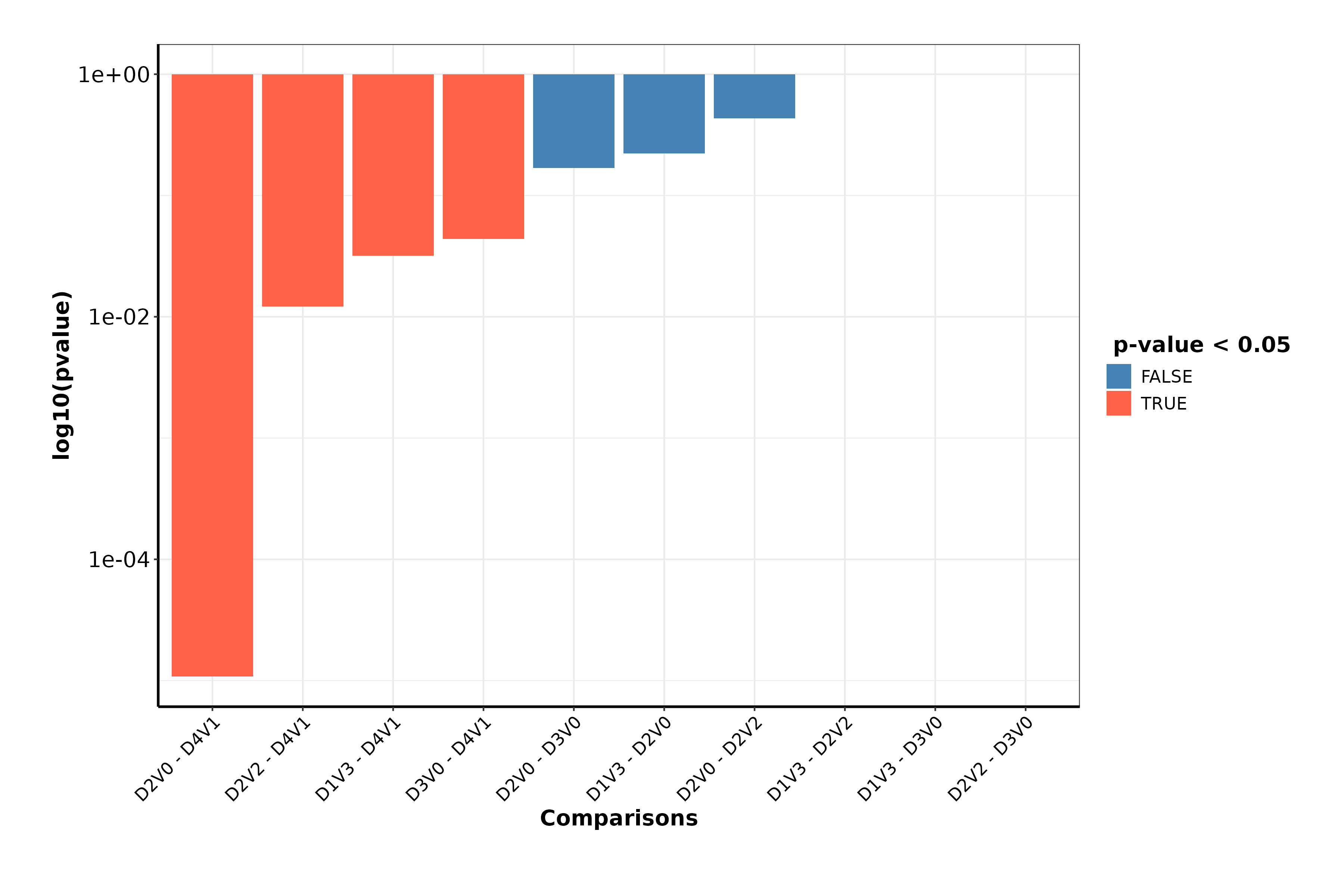}
 \captionsetup{labelformat=empty} % removes "Figure X:"
 \caption{\textbf{Supplementary Figure S7.} \textbf{Results of Dunn's Test p-values for pairwise comparisons between different scenarios for the total number of depopulated animals.} The y-axis represents the negative log-transformed p-values ($\text{-log}_{10}(\text{p-value})$), with red indicating lower p-values (higher significance) with p-value $<$ 0.05.}
 \label{fig:supp_figure_s7}
\end{figure*}
\clearpage
%\subsection*{References}

%% file: table_trasnmission.tex
\begin{table}[H]
\centering
\caption{Host-to-host transmission coefficients ($\beta$) per animal$^{-1}$ day$^{-1}$.}
\label{tab:transmission_beta}
\begin{tabular}{p{5cm}p{5cm}p{5cm}}
\toprule
\textbf{Species or interaction} & \textbf{Value distribution (min, mode, max)} & \textbf{Reference} \\
\midrule

Bovine to Bovine & PERT (0.18, 0.24, 0.56) &  \citep{da_costa_assessing_2022}  \\
Bovine to Swine & PERT (0.18, 0.24, 0.56) & Assumed \\
Bovine to Small ruminants & PERT (0.18, 0.24, 0.56) & Assumed \\
Swine to Bovine & PERT (3.7, 6.14, 10.06) & Assumed; \citep{orsel_quantification_2007} \\
Swine to Swine & PERT (3.7, 6.14, 10.06) & \citep{eble_quantification_2006} \\
Swine to Small ruminants & PERT (3.7, 6.14, 10.06) & Assumed; \citep{orsel_quantification_2007}  \\
Small ruminants to Bovine & PERT (0.044, 0.105, 0.253) & Assumed;  \citep{orsel_quantification_2007}\\
Small ruminants to Swine & PERT (0.006, 0.024, 0.09) &  \citep{goris_quantifying_2009}  \\
Small ruminants to Small ruminants & PERT (0.044, 0.105, 0.253) &  \citep{orsel_quantification_2007}  \\

\bottomrule
\end{tabular}
\end{table}

%% file: infec_incub_periods.tex
\begin{table}[H]
\centering
\caption{Within-farm latent and infectious period distributions.}
\label{tab:S_periods}
\begin{tabular}{p{5cm}p{5cm}p{5cm}}
\toprule
\textbf{Parameter and species} & \textbf{Mean,
median (25th, 75th percentile)} & \textbf{Reference} \\
\midrule

Latent period ($\sigma$), Bovine & 3.6, 3 (2, 5) & \citep{mardones_parameterization_2010} \\
Latent period ($\sigma$), Swine & 3.1, 2 (2, 4) & \citep{mardones_parameterization_2010} \\
Latent period ($\sigma$), Small ruminants & 4.8, 5 (3, 6) & \citep{mardones_parameterization_2010} \\

Infectious period ($\gamma$), Bovine & 4.4, 4 (3, 6) & \citep{mardones_parameterization_2010} \\
Infectious period ($\gamma$), Swine & 5.7, 5 (5, 6) & \citep{mardones_parameterization_2010} \\
Infectious period ($\gamma$), Small ruminants & 3.3, 3 (2, 4) & \citep{mardones_parameterization_2010} \\

\bottomrule
\multicolumn{3}{l}{\textit{Note: All time units are in days.}}
\end{tabular}
\end{table}

%% file: Table1.tex
%\begin{landscape}

\begin{table}[H]
\centering
\caption{Parameters of the model by scenario used in the control action simulations.}
\label{tab:model_parameters}
\footnotesize
\setlength{\tabcolsep}{4pt}
\begin{threeparttable}
\begin{tabular}{@{}lcccccc@{}}
\toprule
\textbf{Parameter} & \textbf{D0V4} & \textbf{D1V3} & \textbf{D2V2} & \textbf{D2V0} & \textbf{D3V0} & \textbf{D4V1} \\
\midrule
\multicolumn{7}{l}{\textit{Initial Conditions}} \\
Infected farm & random & random & random & random & random & random \\
The initial simulation day & random & random & random & random & random & random \\
\midrule
\multicolumn{7}{l}{\textit{Detection}} \\
Days maximum of control action & 120 & 120 & 120 & 120 & 120 & 120 \\
\midrule
\multicolumn{7}{l}{\textit{Control zone(s) radii in kilometers}} \\
\quad Infected zone & 3 & 3 & 3 & 3 & 3 & 3 \\
\quad Buffer zone & 5 & 5 & 5 & 5 & 5 & 5 \\
\quad Surveillance zone & 7 & 7 & 7 & 7 & 7 & 7 \\
\midrule
\multicolumn{7}{l}{\textit{Movement restriction}} \\
Standstill (days) & 30 & 30 & 30 & 30 & 30 & 30 \\
Standstill in infected zone & F & F & F & F & T & F \\
Standstill in buffer zone & F & F & F & F & T & F \\
Standstill in surveillance zone & T & T & T & T & T & T \\
Traceback steps in the contact network & 1 & 1 & 1 & 1 & 2 & 1 \\
\midrule
\multicolumn{7}{l}{\textit{Depopulation}} \\
Farms depopulated per day (limit) & 0 & 1--2 & 2--3 & 2--3 & 3--4 & 6--8 \\
Depopulation in infected zone & NA & F & F & F & F & NA \\
Only depopulate infected farms & NA & T & T & T & T & NA \\
\midrule
\multicolumn{7}{l}{\textit{Vaccination}} \\
Days to achieve immunity & 15 & 15 & 15 & NA & NA & 15 \\
Farms vaccinated in buffer zone/day & 30--50 & 25--45 & 15--30 & NA & NA & 15--20 \\
Farms vaccinated in infected zone/day & 30--50 & 25--45 & 15--30 & NA & NA & 15--20 \\
Vaccine efficacy & 0.8, 0.9, 1 & 0.8, 0.9, 1 & 0.8, 0.9, 1 & NA & NA & 0.8, 0.9, 1 \\
Vaccination of swine & F & F & F & NA & NA & F \\
Vaccination of bovines & T & T & T & NA & NA & T \\
Vaccination of small ruminants & F & F & F & NA & NA & F \\
Vaccination in infected zone & T & T & T & NA & NA & T \\
Vaccination in buffer zone & F & T & T & NA & NA & F \\
Vaccination delay (days) & 15 & 15 & 15 & NA & NA & 15 \\
Vaccination in infectious farms & T & F & T & NA & NA & T \\
\bottomrule
\end{tabular}
\begin{tablenotes}
\footnotesize
\item ``T'' denotes True, indicating that the control action parameter is applied, and ``F'' is false when control actions were not applied. ``NA'' signifies Not Applicable, meaning the parameter configuration is not relevant to the current scenario.
\end{tablenotes}
\end{threeparttable}
\end{table}
%\end{landscape}

%% file: Table2.tex
\begin{table}[H]
\centering
\caption{Performance metrics for disease control simulation scenarios, organized by the percentage of controlled outbreaks. The parameters include the percentage of outbreaks controlled, the duration in days of control actions, the number of vaccinated animals, the number of depopulated animals, and the total infected farms, with each parameter showing median, interquartile range, and maximum values.}
\label{tab:disease_control}
\footnotesize % Smaller font size
\setlength{\tabcolsep}{4pt} % Reduce column spacing
\begin{tabular}{@{}lccccc@{}}
\toprule
\textbf{Scenario} & 
\textbf{Controlled} & 
\textbf{Total Infected} & 
\textbf{Days on} & 
\textbf{Vaccinated} & 
\textbf{Depopulated} \\
& \textbf{Outbreaks} & 
\textbf{Farms} & 
\textbf{Control} & 
\textbf{Animals} & 
\textbf{Animals} \\
& \textbf{(\%)} & 
\textbf{(med, IQR, max)} & 
\textbf{(med, IQR, max)} & 
\textbf{(med, IQR, max)} & 
\textbf{(med, IQR, max)} \\
\midrule
D0V4 & 2.22 & 
\begin{tabular}[t]{@{}c@{}}17 \\ (9.5--37.5, \\ 80)\end{tabular} & 
\begin{tabular}[t]{@{}c@{}}26 \\ (23--52, \\ 71)\end{tabular} & 
\begin{tabular}[t]{@{}c@{}}596,530 \\ (350,817--\\ 938,389, \\ 3,844,170)\end{tabular} & 
\begin{tabular}[t]{@{}c@{}}0 \\ --- \\ ~\end{tabular} \\
\addlinespace
D2V2 & 90.52 & 
\begin{tabular}[t]{@{}c@{}}54 \\ (36--76, \\ 187)\end{tabular} & 
\begin{tabular}[t]{@{}c@{}}36 \\ (22--53, \\ 95)\end{tabular} & 
\begin{tabular}[t]{@{}c@{}}447,278 \\ (258,465--\\ 680,058, \\ 2,711,163)\end{tabular} & 
\begin{tabular}[t]{@{}c@{}}30,894 \\ (18,383--\\ 46,762, \\ 317,969)\end{tabular} \\
\addlinespace
D1V3 & 90.96 & 
\begin{tabular}[t]{@{}c@{}}54 \\ (36--77, \\ 195)\end{tabular} & 
\begin{tabular}[t]{@{}c@{}}35 \\ (22--53, \\ 96)\end{tabular} & 
\begin{tabular}[t]{@{}c@{}}434,240 \\ (256,107--\\ 679,540, \\ 2,685,043)\end{tabular} & 
\begin{tabular}[t]{@{}c@{}}30,444 \\ (18,107--\\ 48,384, \\ 368,609)\end{tabular} \\
\addlinespace
D2V0 & 96.60 & 
\begin{tabular}[t]{@{}c@{}}53 \\ (33--85, \\ 273)\end{tabular} & 
\begin{tabular}[t]{@{}c@{}}21 \\ (12--35, \\ 96)\end{tabular} & 
\begin{tabular}[t]{@{}c@{}}0 \\ --- \\ ~\end{tabular} & 
\begin{tabular}[t]{@{}c@{}}33,097 \\ (19,387--\\ 53,072, \\ 566,938)\end{tabular} \\
\addlinespace
D3V0 & 99.90 & 
\begin{tabular}[t]{@{}c@{}}48 \\ (30--75, \\ 286)\end{tabular} & 
\begin{tabular}[t]{@{}c@{}}14 \\ (8--22, \\ 95)\end{tabular} & 
\begin{tabular}[t]{@{}c@{}}0 \\ --- \\ ~\end{tabular} & 
\begin{tabular}[t]{@{}c@{}}31,467 \\ (18,393--\\ 50,620, \\ 504,437)\end{tabular} \\
\addlinespace
D4V1 & 100 & 
\begin{tabular}[t]{@{}c@{}}41 \\ (27--61, \\ 206)\end{tabular} & 
\begin{tabular}[t]{@{}c@{}}6 \\ (4--9, \\ 41)\end{tabular} & 
\begin{tabular}[t]{@{}c@{}}211,002 \\ (94,520--\\ 378,256, \\ 1,127,699)\end{tabular} & 
\begin{tabular}[t]{@{}c@{}}28,976 \\ (16,836--\\ 45,627, \\ 433,797)\end{tabular} \\
\bottomrule
\end{tabular}
\end{table}